# Ionization cooled ultra pure beta-beams for long distance $\nu_e \to \nu_\mu$ transitions, $\theta_{13}$-phase and CP-violation.

C. Rubbia

Physics Department and INFN, Sezione di Pavia,

Pavia, Italy

**Abstract.**

The key process is the observation of tiny oscillation mixing between $\nu_\mu$ and $\nu_e$ related to so far unknown $\theta_{13}$ amplitude, at distances corresponding to an invariant neutrino flight path around about 2.0 MeV/km. Zucchelli [12] has proposed the production of very pure $\nu_e$ beams (beta beams), in which relativistic radio-nuclides are stored in a high energy storage ring and decay in a long straight section pointing toward the neutrino detector far away. This method produces ultra pure anti-$\nu_e$ (He-6) and $\nu_e$ (Ne-18) with a negligible $\nu_\mu$-contamination ($\leq 10^{-5}$).

A novel kind of beta element production has been recently proposed [15] in which slow (v ≈ 0.1 c) fully ionized ions are stored in a very small storage ring. Products like the isomeric doublet Li-7(d,p) Li-8 and Li-6(He3,n) B-8 are stopped in some foils few µm thick, from which they quickly diffuse as neutral atoms.

The improvements are shown to be considerable. An ion source, a chain of several accelerators in cascade and a high energy storage/decay ring can produce a B-8 induced $\nu_e$ spectrum which has as much as 7.7 times higher energy than the one from He-6 induced anti-$\nu_e$ for a given magnetic rigidity of the storage ring. At a given neutrino energy, the CC cross sections for $\nu_e$ are about 3 x larger than the ones of anti-$\nu_e$ while the NC/CC inelastic pion background faking muons is ≈ 3 x smaller.

An optimal $\nu_e$ source might be fully ionized B-8's from the 120 GeV Main Energy Injector at FNAL, followed by a storage and a neutrino fly path of ≈ 700-800 km. As detector we consider a LAr TPC with a fiducial mass of 50 ÷ 100 kton. Such a technology should permit to detect a background free signal for $\sin^2(2\theta_{13})$ as small as ≈ 6.0 x $10^{-4}$. By comparison, the present experimental limit is < 0.14.

## 1.— Overview.

As well known, crucial but very difficult experiments are needed in order to complete the phenomenology of the neutrino sector. Recently many experiments have provided strong evidence that neutrinos undergo flavour changing transitions [1-7]. Such a wide programme demands new accelerated neutrino beams with well identified initial species and with a long distance between beam and detector. Several other proposals have more conservatively chosen a very massive water Cherenkov detector [9] or a finely segmented liquid scintillator [10]. A massive detector based on the liquid Argon technology [8], provided it can be made massive enough, is probably offering the best opportunities for such future programmes.

Cosmological arguments have suggested that in order to build up the today's dominance of matter over anti-matter in non-equilibrium conditions a strong CP violation in the quark sector must be extended also to the leptonic sector. To this effect, all the three neutrino mixing angles must have non zero values, including the presently unknown $\theta_{13}$, for which the CHOOZ experiment [4] has given the limit $\sin^2(2\theta_{13}) \leq 0.14$ (0.18). As well known, the experimental determination of this phase should complete the already known large angles parameters $\theta_{12}$ and $\theta_{23}$ and the two mass differences $\Delta m_{12}^2$ and $\Delta m_{23}^2$. Provided $\theta_{13} \neq 0$, the leptonic CP violating phase $\delta$ becomes accessible and with sufficient statistics and in absence of backgrounds one might discover both of the CP violating phase $\delta$ and the phase $\theta_{13}$ (Figure 1).

Starting either from an initial high purity $\nu_\mu$ or from a $\nu_e$ source, the key physical process is the observation of tiny $O(\approx 10^{-3})$ oscillation mixing between $\nu_\mu \leftrightarrow \nu_e$ related to $\theta_{13}$ around a distance L centred to a rotation angle $\phi = \pi/2$, where $\phi = 1.27 \Delta m_{23}^2 (eV^2) L(km)/E(GeV)$. This oscillation parameter has been recently determined by SuperKamiokande (SK) [1] to be $\Delta m_{23}^2 = (2.5 \pm 0.5) \times 10^{-3} eV^2$ (90% confidence limit), corresponding to a neutrino flight path of (2.0 ± 0.40) MeV/km for $\phi = \pi/2$. This result has been confirmed by K2K [2].

The neutrino energy domain is therefore directly proportional to the distance between the source and the decay locations, independently of the method chosen to produce the neutrino. The $\phi = \pi/2$ value is $1470 \pm 295$ MeV for the LNGS and Soudan neutrino laboratories (L = 730 km), $434 \pm 120$ MeV for the SK experiment (L = 295 km), and $260 \pm 55$ MeV for the laboratory of Frejus from CERN (L = 130 km).

Since the massive neutrino detector is located at a fixed distance, a wide neutrino energy spectrum must be observed in order to cover different rotation angles. A sizeable event rate must persist over successive maxima of $\phi = (2n-1)\pi/2$, with n = 1, 2, 3, and so on. Indeed the most significant differences in $(\delta, \theta_{13})$ values occur for n > 1. Therefore the full accessibility to an extended neutrino energy region is highly relevant for a comprehensive search of CP-violation. The oscillation mixing probability for different values of the CP violating phase and an arbitrarily chosen value $\sin^2(2\theta_{13}) = 0.012$ is shown in Figure 1 as a function of the reduced distance in MeV/km.

Longer paths are also under consideration, for instance the BNL to Homestake (L = 2540 km) and the FNAL to Homestake (L = 1290 km) configurations. However for these longer configurations, matter oscillation effects become dominant and alter the $\theta_{13}$ oscillation pattern, introducing for instance large differences between initial $\nu$ and $\bar{\nu}$. Therefore distances of the order of about 1000 km represent an acceptable compromise between decay distance and effects due to matter oscillations and both LNGS/CERN and FNAL/Soudan (L = 730 km) or eventually FNAL/Homestake (L= 1290 km) represent optimal choices. However, the ordering of neutrino mass states can also be determined if electron induced matter oscillations propagate to sufficiently long distances (a few thousand km) through the earth.

The present upper limit of $\sin^2(2\theta_{13})$ is 0.14 [4]. Conventional horns and low energy (30-100 GeV) proton beams in MWatt region may be capable of pushing the sensitivity up to $\sin^2(2\theta_{13})$ of maybe $\approx 0.02$, the limit due to the tiny natural $\nu_e$ contamination of the horn driven $\nu_\mu$ beam. Entirely new methods are required if $\sin^2(2\theta_{13})$ would turn out to have an even smaller value.

## 2.— The Zucchelli/CERN proposal.

Considerable interest has arisen the method proposed by Zucchelli [12] of pure $\nu_e, \bar{\nu}_e$ beams from a $\beta$-decay in a decay channel in which relativistic radio-nuclides are stored in a high energy, dedicated storage ring pointing to the far away neutrino detector. In contrast with a conventional beam, the advantage of this method is that very pure $\nu_e$, $\bar{\nu}_e$ beams are produced with a nearly zero initial contamination, $O(\approx 10^{-5})$, for instance caused by the decay daughter nuclei ejected from the beam and interacting with the walls of the storage ring.

The emittance of the stored ion beam in the storage ring is generally small when compared to the one of the β-decay angle. Hence the analysis of the beam is very simple and neutrino fluxes at the detector can be easily determined with the help of the kinematics of the radioactive decay. There is no a priori need for a near-detector. For the radioactive ion beam pointing to the detector, the neutrino energy $E_\nu$ has an end point given by $E_\nu = 2\gamma Q^*$, where $Q^*$ is the beta decay CM end point kinetic energy and $\gamma$ is the relativistic factor of the stored ions. The energy spectrum at the detector is mirroring with a multiplicative accelerating factor the spectrum in the centre of mass of the decaying radio-nuclide. Note that the neutrino average opening angle is $1/\gamma$, independent of the size of $Q^*$. For a given accelerator's $\gamma$ and a given invariant distance $E_\nu/L$ (in MeV/km), the neutrino event rate at the detector is $\propto 1/L$ as resultant of the flux rate's dependence $\propto 1/L^2$ and of the cross section, growing like $E_\nu \propto L$.

Finally, the choice of the best $Q^*$ for the $\nu_e, \bar{\nu}_e$ beams is determined by the availability of a pair of low mass radio-nuclides with an adequate rest half-life — of the order of 1 s — and which can be produced in an sufficiently large amounts in or-der to ensure a high neutrino event rate at hundreds of kilometres away after acceleration and storage.

The Zucchelli/CERN proposal [14] is based on He-6 ($Q^* \approx 3.508$ MeV and $\tau_{1/2}$ = 0.806 s) for $\beta^- \rightarrow \bar{\nu}_e$ and on Ne-18 ($Q^* \approx 4.443$ MeV, $\tau_{1/2}$ =1.67 s) for $\beta^+ \rightarrow \nu_e$. He-6 had been traditionally produced according to the reaction $Be^9(n,\alpha)He^6$, bombarding very finely powdered $Be(OH)_2$ with a high flux of fast neutrons from a nuclear reactor [13]. The fast reactor is here replaced by a high power (MWatts) spallation source driven by a projected superconducting proton high current LINAC of a few GeV [14].

Ne-18 is also produced by high energy neutrons from MgO, but with a rate which is so far estimated to be at least 20 times smaller. Until some novel method is developed in order to produce sufficient amounts of Ne-18, the main thrust is concentrated in the use of He-6.

The He-6 beam [14] is then accelerated in the SPS to an equivalent proton energy of ≈ 300 GeV and accumulated in an additional storage ring with two very long straight sections, at which energy ions spontaneously decay with about 80 s half-life. The He-6 relativistic factor is $\gamma = Z/A \gamma_{proton} = \gamma_{proton}/3 \approx 100$. Because of the low $Q^*$ value of He-6, the end point energy of neutrinos in the laboratory is 700 MeV.

A massive Cherenkov water detector is to be located inside the Frejus tunnel (L = 130 km), where the $\phi = \pi/2$ oscillation angle is $260 \pm 55$ MeV. According to the CERN design parameters, the number of collected He-6 ions in the storage ring will be of the order of 3 x $10^{18}$/year. In order to collect an adequate number of oscillated events, the fiducial mass of a water Cherenkov detector [14] should be about ½ Megaton out of several units, each of 145 kton.

The choice of Frejus and of such a short decay path (L = 130 km) are not without problems. For neutrino energies not too far from threshold, the elastic cross section for complex nuclei is strongly suppressed because of the Pauli factor that requires that the outgoing nucleon should be above the Fermi sea. The fast energy dependence of the cross sections near threshold requires a very accurate energy resolution and seriously complicates the energy inter-calibrations for the $\mu$ and $e$ channels. Individual cross sections are not sufficiently well known and they are presently only roughly estimated theoretically with the help of nuclear physics arguments. They must be experimentally measured with extensive calibration experiments. It is very debatable if the energy dependence of neutrino oscillation pattern could be ultimately determined with a sufficient reliability so very near to the threshold.

The muon-oscillated low energy events are not well separated from the large background of muon events due to atmospheric neutrinos. In order to remove such a background, the beam stored in the ring must be very tightly bunched, with a duty cycle factor that in the CERN design exceeds a factor of 1/1000. This introduces additional limitations to the parameters of the storage ring and to the preceding chain of the many successive particle accelerators.

The choice of a $\nu_e$ instead of the usual $\nu_\mu$ with focusing horn introduces also different backgrounds due to neutral currents events ($NC/CC \approx 0.3$ for $\bar{\nu}_e$) which may generate a $\pi^+$ for instance via the reaction $\bar{\nu}_e + p \rightarrow \bar{\nu}_e + \Delta^+ \rightarrow \bar{\nu}_e + n + \pi^+$. In the water Cherenkov detector considered for the Frejus Laboratory, $\pi^+$ which have not interacted and $\mu^+$ particles have nearly identical signatures. A possible kinematical difference may be associated with the fact that the $\mu^+$ converted from the $\bar{\nu}_e$ is "leading" and therefore has generally an energy quite close to the one of the neutrino, the $\pi^+$ is the result of target fragmentation and therefore has generally smaller energies, slowly dependent of the neutrino energy. But in the case of a $\phi = \pi/2$ oscillation angle for 260 MeV neutrinos, such separation may not be very significant and the required background rejection with the help of kinematics cuts is, in our view, unproven at this stage.

On the other hand, still using He-6 but for instance at a more acceptable distance of L = 730 km, the optimal equivalent proton energy is 730 km/130 km = 5.6 times larger, namely 1.7 TeV, corresponding to He-6 relativistic boost $\gamma_{He-6} \approx 560$. Both accelerator and storage ring become evidently much larger and correspondingly more expensive. The flux at the experiment $\propto \gamma^2/L^2$, taking into account of the longer optimal distance and of the corresponding larger value of $\gamma$, will remain unchanged, but the corresponding neutrino event rate will grow linearly with the energy.

### 3.— Nuclear reaction enhancement with ionization cooling.

The aim of the present paper is the one of looking for other low A radioactive nuclei with the highest possible CM decay energies and a half-life comparable to the one of He-6. C. Rubbia *et al.* [15] have recently discussed a new method of producing and accumulating an adequate number of several radioactive nuclei with low energy two body reactions. Alternatives of radioactive nuclei other than He-8 and Ne-18 become therefore possible.

In particular we have examined the case of the isomeric doublet Li-8 ($\tau_{1/2}$ = 0.84 s) and B-8 ($\tau_{1/2}$ = 0.77 s) (Figure 2) with the reactions Li-7(d,p) Li-8 or Li-6(He3,n) B-8. B-8 is the neutrino spectrum well studied as the main source of high energy Solar Neutrino. It has a complicated decay scheme already described by Bahcall *et al.* [16]: the initial 2+ state of B-8 decays $\beta^+$ in 96 % of the times with 17.979 MeV to a very broad 2+ state at 3.04 MeV, which in turn decays into $2\alpha$ fast break up of the Be-8. Therefore the neutrino energy distribution is substantially different than one of an allowed $\beta$-spectrum with an end point around 15 MeV and it has to be numerically calculated [16]. The neutrino spectra emitted with B-8 and Li-8 are shown in Figure 3, compared to the one of He-6. They are very similar, since the differences are only presumably related to Coulomb corrections, which are very small at such a low A.

This novel type of ionization "cooling" is generally applicable to slow ($v \approx 0.1c$) ions stored in a very small storage ring (Figure 4). The many traversals through a thin foil enhance the nuclear reaction probability, in a steady configuration in which ionisation losses are recovered at each turn by a RF-cavity. For a uniform target "foil", typically few hundred $\mu g/cm^2$ thick, transverse betatron oscillations are "cooled", while the longitudinal momentum spread diverges exponentially since — in the case of $dE/dx$ — faster (slower) particles ionise less (more) than the average. In order to "cool" also longitudinally, a chromaticity has to be introduced with a wedge shaped "foil", such as to instead increase (decrease) the ionisation losses for faster (slower) particles. Multiple scattering and straggling are then "cooled" in all three dimensions, with a method similar to the one of synchrotron cooling, but valid for low energy ions. Particles then stably circulate in the beam, until they undergo for instance nuclear processes in the thin target foil.

The method is suited for the nuclear production of a few MeV/A ion beams. Simple reactions — for instance Li-7(d,p) Li-8 or Li-6(He3,n) B-8 — are preferably produced in the "mirror" kinematical frame, namely with a heavier ion colliding against a gas-jet target [15]. Kinematics is generally very favourable, with angles in a narrow angular cone (around $\approx$ 10 degrees for the mentioned reactions) and a low range outgoing energy spectrum which allows an efficient collection as a neutral gas at high temperature in tiny foils with the help of a technology perfected at CERN-ISOLDE [18]. The neutral gas is then converted with an ion source and accelerated through the appropriate chain of accelerators. This method appears capable of producing a tiny "table top" storage ring with an accumulation rate in the range $10^{13} \div 10^{14}$ ions/sec.

### 4.— An accelerator/storage setup based on β-decays of B-8 and Li-8.

With the choice of B-8 two main factors are reducing substantially the required proton equivalent energy of the accelerator with respect to the Zucchelli/CERN proposal, namely (1) the much higher $Q^*$, which gives an average CM neutrino energy of 7.0 MeV rather than 1.7 MeV (multiplying factor 4.11, see Figure 3) and (2) the larger ratio Z/A = 5/8 = 0.625 rather than 2/6 = 0.333 for He-6, incrementing the value of $\gamma$ by a factor as large as 1.87. Therefore for a given magnetic rigidity or proton equivalent momentum, the choice of B-8 produces neutrinos with an average energy 4.11 x 1.87 = 7.7 times larger !

As a consequence, for instance the *existing* Main Energy Injector at FNAL with 120 GeV protons may be modified in order to produce fully stripped B-8 beta neutrinos with an end point of 2.5 GeV, perfectly suited for L = 730 km and the Sudan mine. The relativistic factor is $\gamma_{B-8} = Z/A \gamma_{proton} = 0.625 \gamma_{proton} = 80$ for the nominal magnetic rigidity. We assume that the improved accelerator complex, now being currently improved to accelerate up to a 2 MWatt proton beam, may be also able to accelerate the same circulating current also for B-8, corresponding to $2 \div 3 \times 10^{13}$ ions/s. Similar modifications may be at hand at CERN in order to produce a sufficiently large B-8 circulating current and a proton equivalent energy in the interval $100 \div 200$ GeV in order to send neutrinos to LNGS laboratory.

The high energy ion beam extracted from the main accelerator is accumulated on a (idealized) storage ring with about 3.14 km circumference and one long straight section pointing to the neutrino detector, comprehending 1/3 of the circumference. The bending radius is assumed to be 3 Tesla. An interesting possibility to be investigated for the storage ring is the relocation of the main magnetic components of the Tevatron, which is expected to terminate its active life around 2009.

Some of the main parameters of the accelerator, the storage ring and the detector are listed in Table 1. The detector distance is 730 km, corresponding to the Soudan or LNGS location. The signal is the neutrino induced $\mu^-$ production in the elastic and inelastic channels. Since the value of $\sin^2(2\theta_{13})$ is presently unknown, five possible values have been listed in Table 2, from $\sin^2(2\theta_{13}) = 0.04$ to $\sin^2(2\theta_{13}) = 0$, namely to the residual rate of the oscillated $\nu_e \rightarrow \nu_\mu$ spectrum coming from the other known amplitudes. As indicative, we have considered a LAr detector of a fiducial mass of 50 kton (35'000 m$^3$) if $\sin^2(2\theta_{13}) > 2.69 \times 10^{-3}$, doubled to 100 kton for an ultimate experimental sensitivity as small as $\approx 6 \times 10^{-4}$. Rates before cuts refer to a 5 years exposure with 200 days/y. For simplicity, the value of the CP violating phase $\delta$ has been set to zero. The dependence of $\delta$ on the oscillation mixing probability has been already described in Figure 1.

The $\nu_e \rightarrow \nu_\mu$ energy spectrum for several different values of $\sin^2(2\theta_{13})$ and $\delta_{CP} = 0$ is shown in Figure 5. As reference it is shown also the case of $\sin^2(2\theta_{13}) = 0$ namely the residual rate of the oscillated $\nu_e \rightarrow \nu_\mu$ spectrum due only to the presence of the other amplitudes. The corresponding ultimate experimental sensitivity to the $\sin^2(2\theta_{13})$ amplitude may be as small as $\approx 6 \times 10^{-4}$.

In Table 2 we show the integrated number of events detected before cuts. The ratio to the number under no oscillation hypothesis (i.e. assuming no neutrino mixing) goes from 2.3 % for $\sin^2(2\theta_{13}) = 0.04$ to 0.18 % for $\sin^2(2\theta_{13}) \approx 6 \times 10^{-4}$.

The main competing backgrounds are here characterized by the neutral current inelastic processes leading to a negative pion capture, $\nu + n \rightarrow \nu + \pi^- + (p)$ and a positive pion decay $\nu + (n,p) \rightarrow \nu + \pi^+ + X$ with the pion misidentified into a negative decaying muon. In view of the large mass of the detector we may safely assume that all charged particles stop within the sensitive volume.

The most effective rejection of the background is related to selecting only the $\mu^-$ signal leading in the LAr TPC to a muon capture (70% of all cases). The pion background is then rejected on the basis of the following criteria:
- The neutral currents background has an experimentally observable rate of the order of 1/60 of the CC current signal without oscillations, although with substantial uncertainties due to the poor experimental spectral information available in the literature.
- With a LAr detector both the total visible energy deposited along the track as well as the range are accurately measured. As shown in Figure 6, for a given deposited energy, the range of a muon track is longer than the corresponding case of a pion. Calculations using the FLUKA simulation [17] indicate that a separation better than 1/200 is possible with a few percent loss of the muon track. (Figure 6).

- About 70% of the negative muons and all pions will undergo nuclear capture at the end of the range. While the $\pi^-$ related star involves the full pion rest mass into several heavy nuclear prongs, most of the $\mu^-$ related energy is emitted by the invisible neutrino. Discrimination of the local energy (blob) measured with LAr at the end point of the track may offer another powerful discrimination between $\pi^-$ and $\mu^-$. Optimisation of this method is still required: at present we assume provisionally at least a factor 50 in the $\pi^-$ misidentification.

Although still somewhat dependent on the actual values for the above indicated effects, it is expected that the number of surviving background events for the 100 kton LAr should be conservatively of the order of about 0.5 events according to the event rates of Table 2 and therefore well manageable when compared even to the smallest data bins of the data of Figure 5.

Note that while the muon of the signal is a leading particle, the background pion is coming from target fragmentation. Therefore neutral current induced events tend to pile up strongly in the region of small lepton energies, where the expected signal due to good $\nu_e \rightarrow \nu_\mu$ events is very small but most sensitive to the specific physical alternatives (see Figure 5).

In analogy with the original Zucchelli's proposal on He-6, $\bar{\nu}_e$ may also be produced, for instance with Li-7(d,p) Li-8 ($\tau_{1/2}$ = 0.84 s). The Li-8, decaying to the same Be-8 final state, at least in the strong interaction limit (no Coulomb corrections) is nearly identical to B-8. However some tiny corrections are experimentally observed (see Figure 3).

The situation is generally less favourable than the case of B-8 and this for several reasons.

- Because of the smaller charge of Li-8 when compared to B-8, the main Energy Injector at FNAL (120 GeV protons) has however a much smaller the relativistic factor $\gamma_{Li-8} = Z/A \gamma_{proton} = 0.375 \gamma_{proton} = 48$, corresponding to a lower $\bar{\nu}_e$ end point of 1536 MeV. Therefore the antineutrino flux at the detector, proportional to $1/\gamma^2$, is only 36 % of the one for B-8.
- The Li-8 CC $\bar{\nu}_e \rightarrow e^+$ cross section without oscillations, averaged over the spectrum is only 11 % of the one of B-8 because of the smaller values of the average cross sections, $\langle \sigma(\bar{\nu}_e \rightarrow e^+) \rangle = 0.16 \times 10^{-38} cm^2$ vs. $\langle \sigma(\nu_e \rightarrow e^-) \rangle = 1.49 \times 10^{-38} cm^2$
- However, for a specified accelerated current, the number of injected ions may be significantly larger because of the fully stripped charge is 3 instead of 5.

The $\bar{\nu}_e$ rates are therefore about a factor 10 smaller than the relevant ones of Figure 5. For instance for 100 kton LAr detector, 5 years of exposure at 200 days/y, the non oscillated antineutrino reference signal is only 18k events, to be compared with the 221 k events of B-8 neutrinos of Table 2. The event distribution is shown in Figure 7 for $\sin^2(2\theta_{13}) = 0.043$.

The event rate may be substantially increased if an additional post-acceleration is provided for instance in the storage ring before decay. As shown in Figure 7, the neutrino rate is quickly rising as soon as $\gamma_{Li-8}$ is increased. The number of oscillated $\overline{\nu}_e \rightarrow \overline{\nu}_\mu$ for $\sin^2(2\theta_{13}) = 0.043$ is rising from 239 events at $\gamma_{Li-8} = 48$ ($E_p = 120$ GeV) to 2274 events at $\gamma_{Li-8} = 80$ ($E_p = 200$ GeV) $= \gamma_{B-8}$ and to 9504 events at $\gamma_{Li-8} = 120$ ($E_p = 300$ GeV).

In the case of $\overline{\nu}_e$ initiated events, the most serious concern is caused by the pion background generated by the neutral currents channel, $\overline{\nu} + (p,n) \rightarrow \overline{\nu} + \pi^+ + X$. As well known, $N(NC)/N(CC) \approx 3/1$ for $\overline{\nu}$'s compared to $\nu$'s. Both $\pi^+$ and $\mu^+$ show the presence of a decay electron, although $\pi^+ \rightarrow \mu^+ + \nu$ is too short to be detectable in the LAr detector. The main surviving criterion of discrimination is the range-energy curve of the relevant track, as shown in Figure 6. It may not be sufficient to discriminate sufficiently $\mu^+$ against $\pi^+$, unless $\sin^2(2\theta_{13})$ is large. It may be however taken into account that the LAr detector is actually segmented in a very large number of separate $dE/dx$ elements, each 3 mm long, and a more sophisticated analysis, taking into account selectively of the individual Landau fluctuations may provide selection criteria more advanced than the simple range-energy sum of Figure 6.

**5.— Conclusions.**

The value of $\sin^2(2\theta_{13})$, the key value to CP violation in the leptonic sector is presently unknown. If $\sin^2(2\theta_{13}) \geq 0.02$, an appropriate $\nu_\mu$ beam with classic horn and high proton intensity (1÷2 MWatt on target, either on-axis or off-axis) has been under consideration in several different Laboratories, since the unwanted, tiny $\nu_e$ beam contamination, typically 0.007 ÷ 0.009 of the number of focussed $\nu_\mu$, is sufficiently small in order to identify the emergence of a $\theta_{13}$ signal. The main background above the intrinsic $\nu_e$ contamination $\nu_e + N \rightarrow e^- + X$ is then due to inelastic neutral currents $\nu + p \rightarrow (p) + \pi^o + \nu$ and $\nu + n \rightarrow (n) + \pi^o + \nu$, with the $\pi^o$ faking the electron. We believe that such a separation is difficult in a large Cherenkov water detector: a more advanced LAr detector is necessary. With a LAr detector, one can promptly distinguish unequivocally the topology of a single electron track from one of the vertex of a $\gamma$ from a $\pi^o$ event.

In a second line of approach, on a longer time scale and especially if it would have turned out that $\sin^2(2\theta_{13}) < 0.02$, entirely new accelerator technologies must be developed, maintaining however, the concept of a very massive LAr detector.

Considerable interest has arisen on the method proposed by Zucchelli of ultra pure $\nu_e, \overline{\nu}_e$ beams from radioactive nuclei produced by a high energy, high power dedicated neutron spallation target. The optimal choice is then $\overline{\nu}_e$ in He-6, with a much lower rate (1/20) of $\nu_e$ in Ne-18.

The recently developed ionization cooling of very intense B-8 and Li-8 beams offers additional interesting possibilities. Compared with He-6 and Ne-18, these decays have comparable half-lives but a much higher beta decay CM kinetic energies. The higher $Q^*$ reflects itself into correspondingly higher neutrino energies. While the optimal choice for He-6 are anti-neutrinos, the best case with ionization cooling are now B-8 neutrino.

The advantages of this new line of approach are considerable. For a given relativistic ion $\gamma$ factor, neutrino energies with B-8 are 4.11 times larger than the ones of He-6. However the relevant factor is the magnetic rigidity of the ring, namely the equivalent proton energy, which depends on the Z and A of the respective ions, which is here $\gamma_{B-8} = 1.87 \gamma_{He-6}$. In conclusion:

(1) for a given proton energy accelerator, the energy spectrum of B-8 is as much as 4.11 x 1.87 = 7.7 times larger than the one of He-6.
(2) The higher relativistic factor, again for a given proton energy, introduces a increase of the beam intensity which is $\propto \gamma^2$, namely a neutrino flux in favour of B-8 of a factor 3.5, which is by no mean negligible.
(3) At a given neutrino energy, the CC cross sections for $\nu_e$ are about 3 x larger than the ones of $\bar{\nu}_e$.
(4) The NC/CC inelastic pion background faking muons is $\approx$ 3 x smaller.
(5) The $\pi/\mu$ selectivity is orders of magnitude more effective with negative tracks from B-8 than with positive tracks from He-6.

With all these factors in mind an existing accelerator complex like the Main Energy Injector at FNAL and 120 GeV protons may be improved in order to produce fully stripped B-8 neutrinos with an end point of 2.5 GeV, namely a range of values in MeV/km which is perfectly suited with an optimal length L = 730 km to the Sudan mine. With an estimated B-8 intensity of 3 x $10^{13}$ ion/s and a LAr detector of 50 ÷ 100 kton (see Table 1), it should be possible to explore comprehensively, with adequate statistics and with a small NC residual background the whole range of possible values of $\sin^2(2\theta_{13})$ to the lowest values compatible with a zero angle and due only to the known presence of the other amplitudes (Figure 5 and Table 2).

It should be underlined however that such a large ion intensity in the Main Energy Injector at FNAL requires a substantial R&D work, in order to ensure that the scenario described in Ref. [15] may be indeed capable of such a remarkable performance and in particular that such an adequate ion accumulation rate can be produced and sustained. Clearly the next step is the experimental realization of a smaller scale test prototype with ionization cooling along the lines of Figure 4.

It should be finally stressed that the complexity and cost of the B-8 ionization cooler is much smaller than the one required for a > 1 GeV high current SC-LINAC and the subsequent MWatt He-6 spallation source and that the consequent factor 8 effectiveness in the magnetic rigidity of the main accelerator and of the storage ring should also be very cost effective.

## 6.— Acknowledgements

The author would like to thank A. Ferrari, J. Kadi and V. Vlachoudis, staff members of CERN-AB Dept. for their valid contributions and for numeric computations on several of the aspects of the subject of this paper. Useful conversations must be acknowledged also with the beta-beam study group [http://beta-beam.web.cern.ch] at CERN and in particular with Mats Lindroos.

Table 1. Main parameters of the B-8 source from the the Main Injector at FNAL

| *Neutrino production* | | | |
|---|---|---|---|
| End point $\beta$-decay energy : | $\approx 15$ | MeV | see Ref. [16] |
| Ion charge: | 5 | | |
| Ion atomic number: | 8 | | |
| Half-life at rest: | 777.00 | ms | |
| Produced B-8 rate: | 3.0 | x$10^{13}$ dec/s | |
| | | | |
| *Accumulator properties* | | | |
| Proton equivalent energy: | 120.9 | GeV | |
| Gamma ion: | 80.000 | | |
| Ion energy: | 600.45 | GeV | |
| Guide field: | 3.0000 | Tesla | |
| Bending radius: | 133.43 | m | |
| Circumference | 3.12 | km | |
| Arc radius: | 166.79 | m | |
| Bending arcs circumf: | 1.048 | km | |
| Decay straight length: | 1.048 | km | |
| Fraction neutrino in axis: | 0.33333 | | |
| Revolution time: | 10.48 | µs | |
| Beam/s current to ring: | 2290.1 | mA | |
| In flight half-life: | 62.16 | sec | |
| Stack current: | 142.35 | A | |
| Neutrino max. energy: | 2713.2 | MeV | |
| | | | |
| *Detector properties* | | | |
| Detector distance: | 730.00 | km | |
| π/2 oscillated energy: | 1781.7 | MeV | |
| Angle of 1 cm at det.: | 0.13699E-01 | µrad | |
| Equiv. Angle CM: | 2.1918 | µrad | |
| Fraction neutrino within 1 cm$^2$ | 0.12743E-12 | | |
| Neutrino flux: | 3.8228 | n/sec/cm$^2$ | |
| Nu-e average cross section: | 1.4965 | $10^{-38}$ cm$^2$ | |
| Detector mass(Ar): | 50.0 | kton | |
| CC Event rate without oscill.: | 149.77 | ev/day | |

Table 2. Beta beam rates for B-8 ions produced with the Main Energy Injector at FNAL and $\gamma_{B-8} = 80$ as a function of the value of $\sin^2(2\theta_{13})$. The LAr detector is at L = 730 km. Nominal exposure: 5 years at 200 days/y.

| $\sin^2(2\theta_{13})$ | 4.30E-2 | 1.10E-2 | 2.69E-3 | 6.71E-4 | 0.00E+0 |
|---|---|---|---|---|---|
| ktons of LAr, fiducial | 50 | 50 | 100 | 100 | 100 |
| All events, no mixing | 110'500 | 110'500 | 221'000 | 221'000 | 221'000 |
| Oscill. events, no cuts | 2276.9 | 715.05 | 704.14 | 557.41 | 556.02 |
| Ratio oscill/all, no mix | 2.06E-2 | 6.47E-3 | 3.19E-3 | 2.52E-3 | 2.52E-3 |

## 7.— References.

## 8.— Figure captions.

Figure 1. Effects of a CP violating amplitude $\delta$ for $\theta_{13} = 3°$ ($\sin^2(2\theta_{13}) = 0.01$). The oscillation mixing probability as a function of the distance in MeV/km for an initial $\mu_e$ into a $\mu^-$ without CP violation ($\delta = 0°$) and for three different values of the CP phase $\delta = 90°, -90°, 180°$.

Figure 2. Isospin triplet with A = 8 (Li-8, Be-8, B-8), decaying to the fundamental level of Be-8. In absence of Coulomb corrections, the three states would have identical nucleons configurations because of charge independence. The actual experimental values of the beta decaying doublet Li-8 with $\tau_{1/2}$ = 0.84 s and B-8 with $\tau_{1/2}$ = 0.77 s are respectively $Q^*$ = 16.005 MeV and $Q^*$ = 16.957 MeV.

Figure 3. Neutrino spectra of Li-8 and B-8, compared to the case of He-6. The two spectra have been calculated with the decay scheme of Figure 2 and following Ref. [16].

Figure 4. Principle diagram [15] of the ionization cooled ion accumulation in a storage ring. Singly charged ions from a low energy accelerator are fully stripped by a thin gas jet target and stably stored inside the storage ring. The energy is continuously lost by the gas target and recovered by a RF cavity, until a nuclear collision ejects the particle from the gas. The produced radioactive ions are collected in the thin reaction stopper and brought to rest. The technique of using very thin targets in order to produce secondary neutral beams has been in use for many years.

Figure 5. Differential spectra (events/100 keV) as a function of the B-8 neutrino energy detected 730 km away (Soudan) from the FNAL Main Energy injector followed by a B-8 decay storage ring (see Table 1). Events rates (see Table 2) are given for 5 years of exposure and 200 days/year and a 50/100 kton fiducial LAr detector. Data are shown for different values of $\sin^2(2\theta_{13})$ to the lowest values compatible with a zero angle and only due to the known presence of the other amplitudes.

Figure 6. Range distribution in cm of LAr for pions and muons for 100 MeV of deposited energy. The calculations take correctly into account all relevant phenomena of energy losses with the help of a FLUKA Monte Carlo programme [17]. It is apparent that in a LAr-TPC pions and muons have distinguishable ranges.

Figure 7 Differential spectra (events/100 keV) for $\sin^2(2\theta_{13})$ = 0.043 as a function of the antineutrino energy for Li-8. The magnetic rigidity of the ring corresponds to $\gamma_{Li-8}$= 48 ($E_p$ =120 GeV). The $\bar{\nu}_e$ rates are about a factor 10 smaller than the ones of Figure 5 for 100 kton LAr detector, 5 years of exposure at 200 days/y, the non oscillated antineutrino reference signal before cuts is only 18k events, to be compared with the 221 k events of B-8 neutrinos.

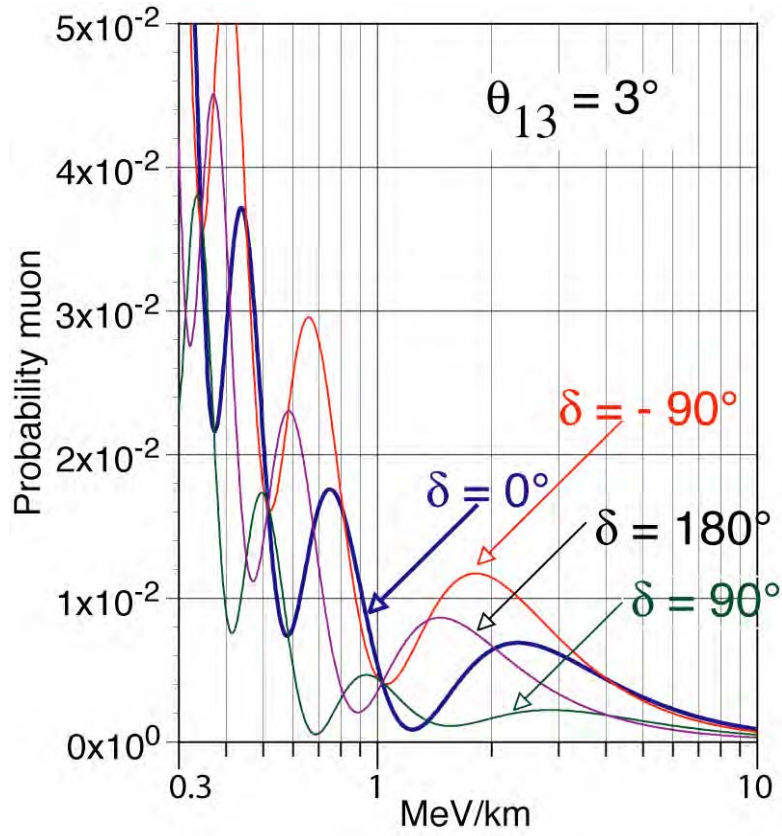

**FIGURE 1.**

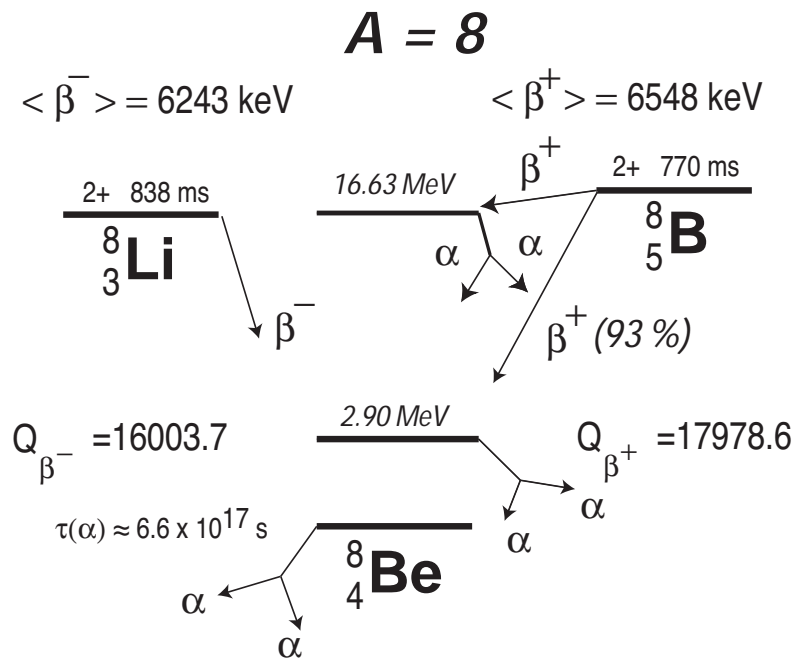

**FIGURE 2.**

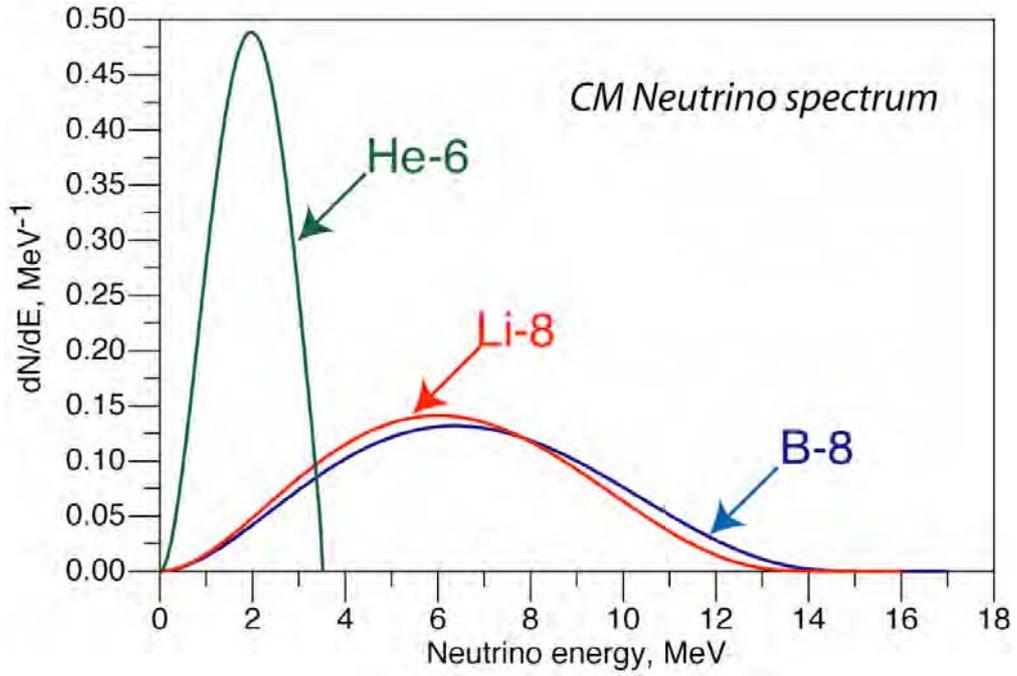

**FIGURE 3.**

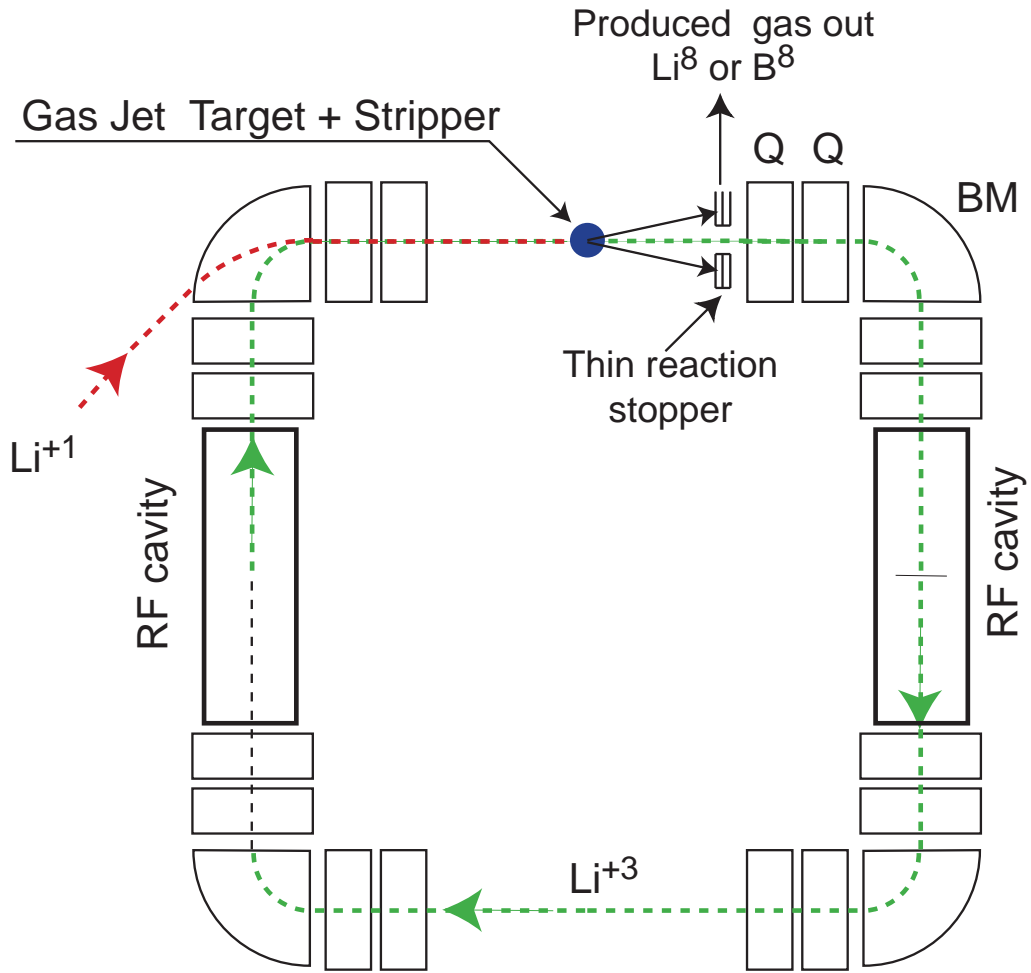

**FIGURE 4.**

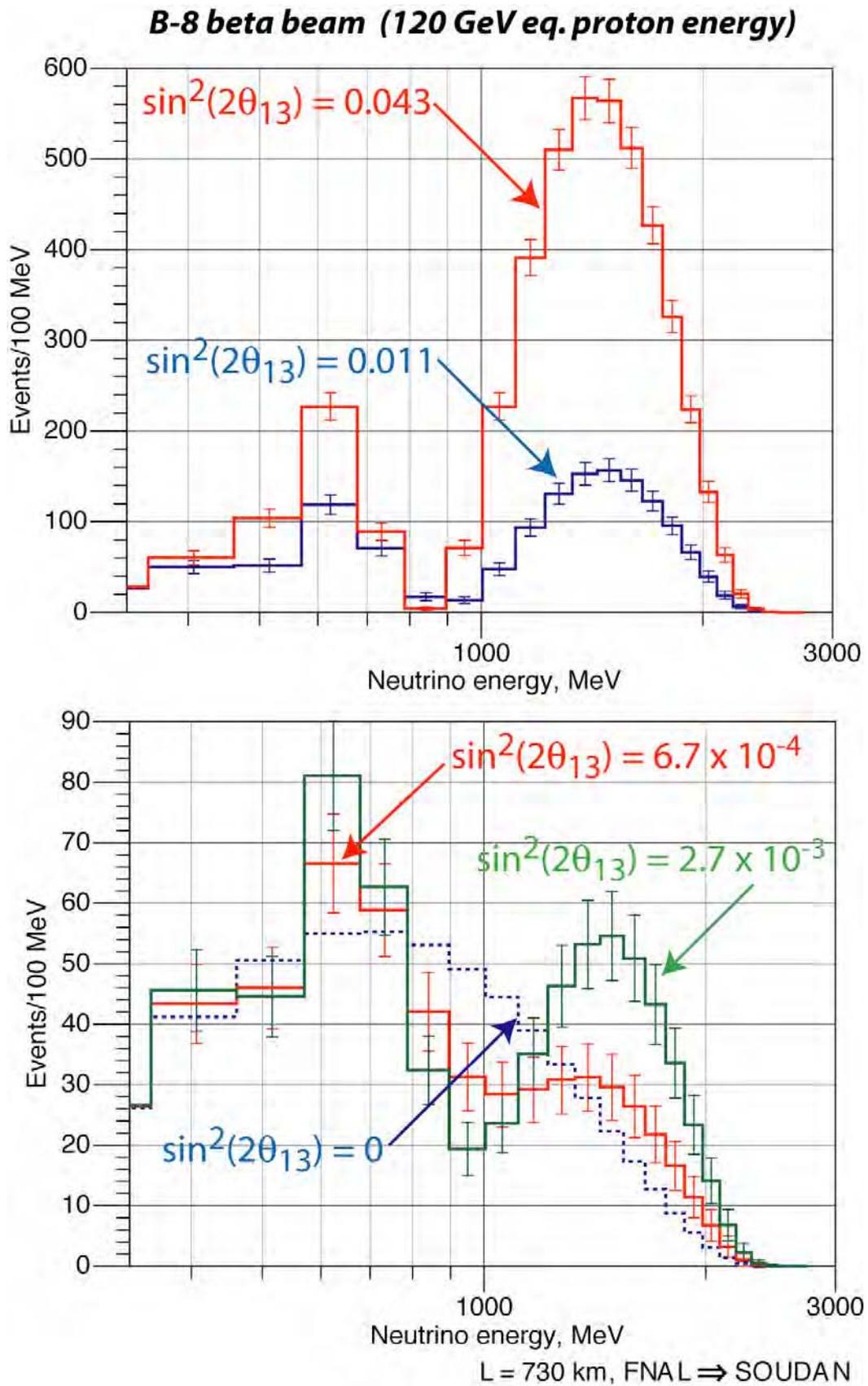

FIGURE 5.

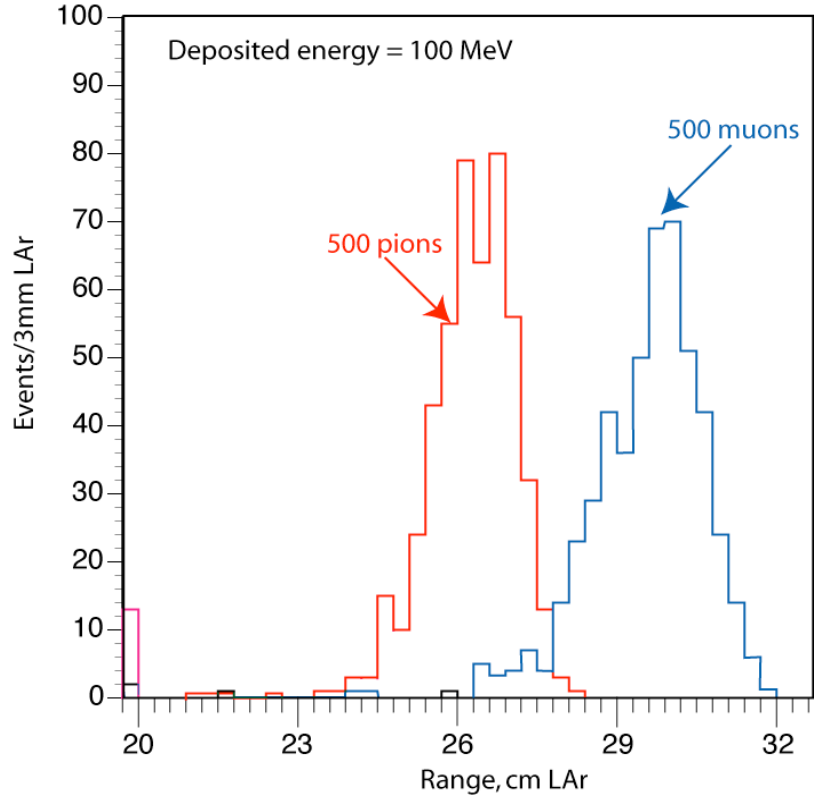

FIGURE 6.

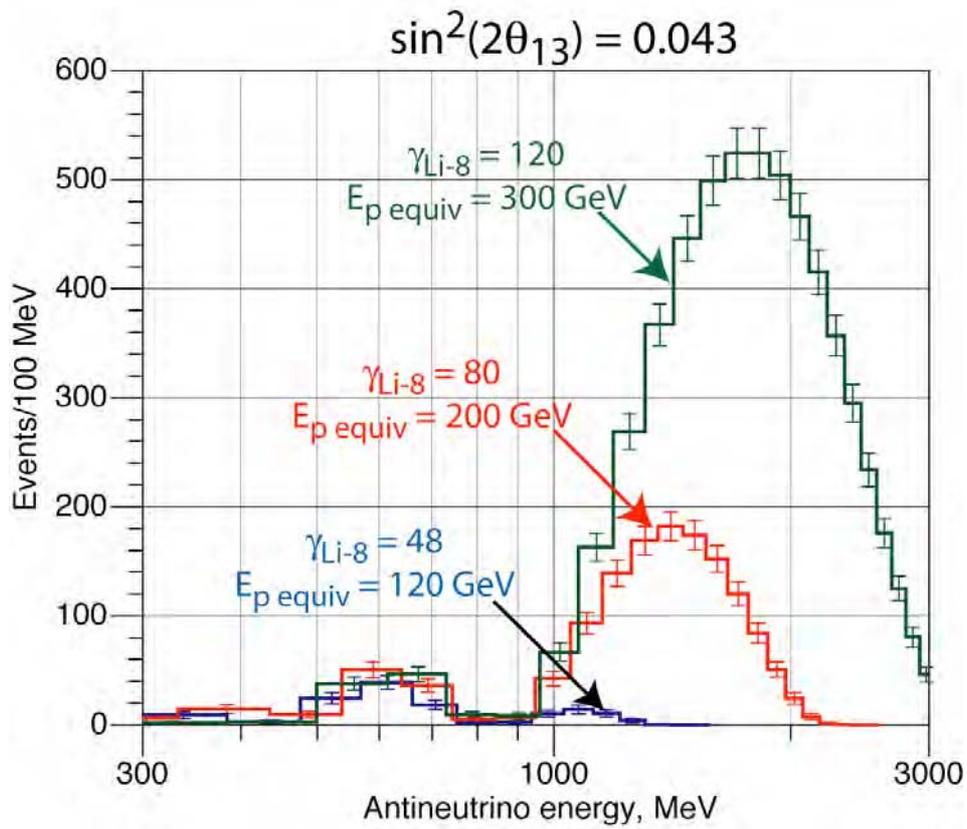

FIGURE 7.